\documentclass[aps,prl,preprint]{revtex4}

\pdfoutput=1
\usepackage[usenames,dvipsnames]{color}
\usepackage{graphicx}
\usepackage{amsmath}
\usepackage{amsfonts}
\usepackage{amssymb}
\usepackage{dsfont}
\usepackage{dcolumn}
\usepackage{bm}
\usepackage{slashed}
\usepackage{pstricks}
\usepackage{slashed}
\usepackage{multirow}
\usepackage{array}
\usepackage{rotating}
\usepackage{xcolor}
\usepackage{epstopdf}
\usepackage{fancybox}
\usepackage{ulem,fancyvrb}

\newcolumntype{x}[1]{
{\centering}p{#1}}%
\newcommand{\tnhl}{\tabularnewline\hline}

\def \cha{\widetilde{\chi}^{\pm}_1}

\newcommand{\beqn}{\begin{eqnarray}}
\newcommand{\eeqn}{\end{eqnarray}}
\newcommand{\be}{\begin{equation}}
\newcommand{\ee}{\end{equation}}

\newcommand{\mathsym}[1]{{}}

\DeclareMathOperator{\sgn}{sgn}

\def \cha{\tilde{\chi}^{\pm}_1}

\newcommand{\na}{\ensuremath{\tilde{\chi}^{0}_1}}
\def \nb{\tilde{\chi}^{0}_2}

\def \n34{\tilde{\chi}^{0}_{3,4}}
\newcommand{\g}{\ensuremath{\tilde{g}}}

\def \ta{\tilde{t}_1}

\def \ba{\tilde{b}_1}

\newcommand{\sta}{\ensuremath{\tilde{\tau}_1}}

\def \smr{\tilde{\mu}_R}
\def \ser{\tilde{e}_R~}

\def\met100{\slashed{E}_T\geq 100 \GeV}
\newcommand{\gappeq}{\mathrel{\rlap {\raise.5ex\hbox{$>$}}
{\lower.5ex\hbox{$\sim$}}}}
\newcommand{\lappeq}{\mathrel{\rlap{\raise.5ex\hbox{$<$}}
{\lower.5ex\hbox{$\sim$}}}}

\def\met{\slashed{E}_{T}}
\def\meff{M_{\rm eff}}

\def\ra{\rightarrow}
\newcommand{\TeV}{\ensuremath{\mathrm{Te\kern -0.1em V}}}
\newcommand{\GeV}{\ensuremath{\mathrm{Ge\kern -0.1em V}}}
\newcommand{\MeV}{\ensuremath{\mathrm{Me\kern -0.1em V}}}

\newcommand{\pb}{\ensuremath{\mathrm{pb}^{-1}}}
\newcommand{\fb}{\ensuremath{\mathrm{fb}^{-1}}}
\newcommand{\bs}{\ensuremath{B_s^0}}

\newcommand{\mm}{\ensuremath{\mu^{+}\mu^{-}}}

\newcommand{\bsmm}{\ensuremath{\bs\ra\mm}}

\newcommand{\brbsmm}{\ensuremath{\mathcal{B}r(\bsmm)}}

\begin{document}
\begin{center}
\end{center}

\title{Excess Observed in CDF $B^0_s \to \mu^{+} \mu^{-}$  and SUSY at the LHC}
\author{Sujeet~Akula$^a$, Daniel~Feldman$^b$, Pran~Nath$^a$ and Gregory~Peim$^a$ }
\affiliation{$a$ Department of Physics, Northeastern University,
 Boston, MA 02115, USA   \\
 $b$ Michigan Center for Theoretical Physics,
University of Michigan, Ann Arbor, MI 48109, USA
 }


%%%%%%%%%%%%%%%%%%%%%%%%%%%%%%%%%%%%%%%%%%%%%%%%%%%%%%%%%%%%%%%%%%%%%%%%%%%%%%%%%%%%%%%%%%%%%%%%%
\begin{abstract}

The recent excess observed by CDF
in $B^0_s \to \mu^{+} \mu^{-}$  is interpreted in  terms of a possible supersymmetric origin.  
An analysis is given of the parameter space of mSUGRA and non-universal SUGRA models under the combined constraints
from LHC-7 with 165~pb$^{-1}$ of integrated luminosity, under the new XENON-100 limits on the 
neutralino-proton spin independent cross section and under the CDF $B^0_s \to \mu^{+} \mu^{-}$
90\% C.L. limit reported to arise from an excess number of dimuon events. It is found that
the predicted value of the branching ratio $B^0_s \to \mu^{+} \mu^{-}$   consistent with all the constraints contains the following set of NLSPs: 
 chargino, stau, stop or CP odd (even) Higgs. 
   The lower bounds of sparticles,
 including those from the LHC, XENON and CDF $B^0_s\to \mu^+\mu^-$ constraint, are exhibited
and the shift  in the allowed range of sparticle masses arising solely due to the extra constraint from the CDF result is given. 
It is pointed out that the two sided CDF 90\% C.L. limit puts upper bounds on sparticle 
masses. An analysis of possible
signatures for early discovery at the LHC  is carried out  corresponding to the signal region in $B^0_s \to \mu^{+} \mu^{-}$.  Implications of GUT-scale non-universalities
in the gaugino and Higgs sectors are discussed. If the excess seen by the CDF Collaboration is supported
by further data from LHCb or D0, this new result could be a harbinger for the discovery of supersymmetry.

\end{abstract}

\maketitle 

\section{I. Introduction}
\begin{figure}[h!]
\begin{center}
\includegraphics[scale=.3]{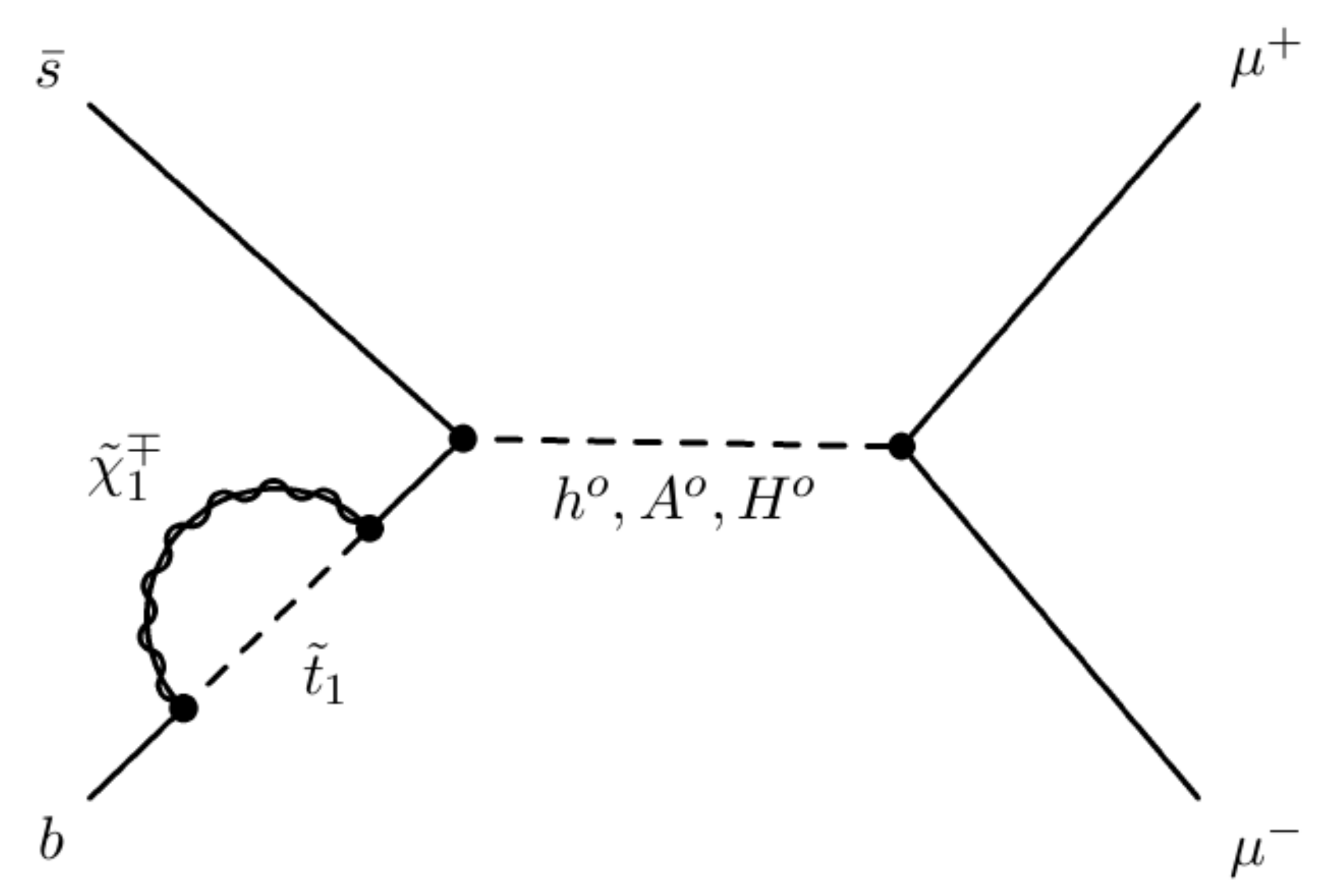}
\end{center}
\caption[]{\label{feyn} Example of  a  diagram giving rise to  supersymmetric contributions  to the process $B_s^0\to \mu^+\mu^-$
producing a scattering amplitude proportional to $\tan^3\beta$.}
\vspace*{-2mm}
\end{figure}

Recently the CDF collaboration has reported an excess in the rare decay $\bsmm$~\cite{Collaboration:2011fi}
using 7~fb$^{-1}$ of integrated luminosity.  Thus CDF Collaboration gives a determination  
\be  
\brbsmm= (1.8^{+1.1}_{-0.9})\times 10^{-8} 
\label{1}
\ee
while the standard model gives $\brbsmm$ to be
$(3.2 \pm 0.2) \times 10^{-9}$~\cite{Buras:2010mh}.
 Supersymmetry proves to be a prime candidate for explaining this excess. The SUSY contribution arises dominantly from the 
Higgs exchange (see Fig.(\ref{feyn})) and this diagram enhances  
$\bsmm$ for large $\tan\beta$  which 
is proportional to $\tan^6\beta$~\cite{Babu:1999hn, Bobeth:2001sq,ddnr,Arnowitt:2002cq,tata}.
$\bsmm$ is also very sensitive to CP violations in the soft sector~\cite{Ibrahim:2002fx,Ibrahim:2007fb},
where it is seen that the effect of CP phases can modify the branching ratio 
by an order of magnitude or more. 
 
 In this work we analyze the implications of these
results in the framework of supergravity, which in the minimal case, mSUGRA~\cite{sugra,hlw},
consists of the parameter space 
\beqn
m_0, ~m_{1/2}, ~A_0, ~\tan\beta, \sgn{\mu},
\label{2}
\eeqn
where $m_0$ is the universal scalar mass, $m_{1/2}$ is the universal gaugino mass, $A_0$ is the
universal trilinear coupling, $\tan\beta$ the ratio of the two Higgs VEVs of MSSM, and $\mu$
is the Higgs mixing parameter. 
The RG analysis of sparticle spectrum of the model
was discussed in~\cite{ArnowittNath}. 
Since the physics at the Planck scale is still largely unknown, inclusion of 
non-universalities in the soft parameters at the unification scale may be desirable.
Such non-universalities must be consistent with flavor changing neutral currents (FCNC).
Two sectors where non-universalities can be introduced consistent with FCNC
are the gaugino mass sector and the Higgs mass sector.
In this work we will first carry out an analysis within the framework of mSUGRA 
and later discuss these non-universal cases.

It is known that the $\brbsmm$ limits have strong implications for 
a variety of SUSY phenomena. Thus the implication of the previous limits on 
$\brbsmm$ regarding the constraints on the
CP odd Higgs~\cite{Feldman:2007fq}, on the neutralino mass~\cite{lowmassneutralino}
and on the spin-independent neutralino-proton cross section~\cite{Feldman:2009pn} 
have been investigated. However, all the
previous analyses used only the upper limit constraint on $\brbsmm$.
 Here we 
investigate the implications arising from the {\it two sided limit} on $\brbsmm$.
Specifically in the analysis of this work we will use the 90\% C.L. limit 
\beqn
4.6\times 10^{-9} < \brbsmm < 3.9 \times 10^{-8}.
\label{3}
\eeqn

The  
effective Hamiltonian governing the decay $\bsmm$ 
is given by~\cite{Bobeth:2001sq,Ibrahim:2002fx} 
\begin{eqnarray}
H_{eff}=-\frac{G_Fe^2}{4\sqrt 2 \pi^2} V_{tb}V_{ts}^* 
\left(C_S O_S + C_P O_P +C_S' O_S'
+ C_P' O_P'+ C_{10} O_{10}\right)_Q.
\end{eqnarray}
Here $O's$  are the effective dimension six operators defined by 
\begin{eqnarray}
O_S= m_b  (\bar s P_R b)(\bar \mu \mu),~~
O_P=m_b (\bar s P_Rb)(\bar \mu\gamma_5 \mu),\nonumber\\
O_S'= m_{s}  (\bar s P_Lb)(\bar \mu \mu),~~
O_P'= m_{s}  (\bar s P_Lb)(\bar \mu \gamma_5 \mu),\nonumber\\
O_{10}=  (\bar s \gamma^{\mu}P_Lb)
(\bar \mu\gamma_{\mu}\gamma_5 \mu),
\end{eqnarray}
where $C's$ are the Wilson co-efficients and $Q$ is the renormalization group  scale.  
The branching ratio $\brbsmm$
 is given by  (see e.g.,~\cite{Ibrahim:2002fx}) 
  \beqn
\brbsmm   &=&
 \frac{G_F^2\alpha^2M_{B_{s}}^5\tau_{B_{s}}}{16\pi^3}
 |V_{tb}V_{ts}^*|^2 \nonumber\\
& \times& \left(1-\frac{4 m_{\mu }^2}{M^2_{B_{s}}}\right)^{1/2}
 \left\{\left(1-\frac{4 m_{\mu }^2}{M^2_{B_{s}}}\right) |f_S|^2+|f_P+2m_{\mu} f_A|^2\right\}.
 \eeqn
Here $f_i$ (i=S,P) and $f_A$ are given by 
 \beqn
 f_i= -\frac{i}{2}f_{B_{s}}\left(\frac{C_im_b-C_i'm_{s}}{m_{s}+m_b}\right),~~~ 
 f_A=-\frac{if_{B_{s}}}{2M_{B_{s}}^2} C_{10}
 \eeqn
 where and $f_{B_s}$ is the decay constant of the $\bs$ meson.
 In addition to the above  there are SUSY QCD effects which 
 have a $\tan\beta$ dependence~\cite{Carena:2000uj}. 
 The Wilson
 co-efficients are  computed in a number of works (see, e.g.,~\cite{Buras:2001mb,Bobeth:2001sq,Ibrahim:2002fx}).
An approximate result for $\brbsmm$ arising from supersymmetry, in the large $\tan \beta$ limit reads~\cite{Buras:2001mb,Bobeth:2001sq} 
\beqn
\label{burasetal}
\brbsmm & \simeq &
3.5\times10^{-5}
\left[\frac{\tau_{B_s}}{1.5 ps}\right]\left[\frac{f_{B_s}}{230\rm{
MeV}}\right]^2
\left[\frac{\left|V_{ts}\right|}{0.040}\right]^2\nonumber\\
&&\times
\left[\frac{\tan\beta}{50}\right]^6\frac{m_t^4}{m_A^4}\frac{(16\pi^2)^2\epsilon_Y^2}{(1+(\epsilon_0 + \epsilon_Y y^2_t)\tan\beta)^2
(1+\epsilon_0\tan\beta)^2}.
\label{4}
 \eeqn 
In the above, 
$V_{ts}$ is a CKM mixing matrix element and
 $\tau_{B_s}$ is the mean lifetime.
Here one can see explicitly the large $\tan \beta$ enhancements.
$\epsilon_0$ and $\epsilon_Y$ are  loop factors given by~\cite{Buras:2001mb}
\beqn
\epsilon_0 & \simeq& -\frac{2 \alpha_s}{3\pi}  \frac{\mu}{m_{\tilde g}} \cdot  H(x^{Q/g}, x^{D/g})\\
\epsilon_Y &\simeq& \frac{1}{16\pi^2} \frac{A_t}{ \mu} \cdot   H(x^{Q/\mu}, x^{U/\mu})\\ 
H(x_1,x_2) &=& \frac{x_1 \ln x_1}{(1-x_1)(x_1-x_2)} + \frac{x_2 \ln x_2}{(1-x_2)(x_2-x_1)}
\label{5}
\eeqn
where $x^{Q/g} = m^2_{Q}/m^2_{\tilde g}, x^{Q/\mu} = m^2_{Q}/\mu^2$, etc.  are written in terms
of the  sbottom and stop masses for $\epsilon_0,\epsilon_Y$ respectively. Analysis is done without
the above approximation, however generically it is a useful guide to the behavior of the $\bs$ branching ratio at large $\tan \beta$.

We will investigate the  constraint of Eq.(\ref{3}) in conjunction with the Large Hadron Collider 
constraint  at $\sqrt{s}=7~\TeV$ (LHC-7)~\cite{cmsREACH,AtlasSUSY,atlas0lep,atlas165pb} as well as the constraint
arising from dark matter direct detection experiments~\cite{cdms,xenon}, on the parameter space of the
mSUGRA model and models with non-universal soft breaking, as well as on the sparticle masses in the framework of the sparticle landscape~\cite{Feldman:2007zn}.
We will also investigate the potential for imminent discovery at LHC-7
of supersymmetric models, which produce a prediction for  $\brbsmm$ that can give rise  to the CDF result. Implications of non-universalities on the 
soft parameters~\cite{gaugino,an,nonuni2,Baer,martin} will also be discussed.  

The outline of the rest of the paper is as follows: In Sec.(II) we discuss  the model space 
investigated in  our survey and describe in detail
the constraints imposed. 
Next, we discuss in a systematic manner the constraints on the remaining parameter space that arise
from LHC-7, XENON-100, and the two-sided limit on $\brbsmm$ from CDF. 
The region of the parameter space consistent with all constraints is then identified. The
mass ranges on some of the lightest sparticles are given showing how these ranges   are modified
by applying the CDF constraint on $\brbsmm$.
In Sec.(III) we discuss LHC signatures for models that may produce the excess in the CDF signal region 
and lie within the discovery reach of LHC-7. In Sec.(IV) we discuss the implication of including the
non-universalities~\cite{gaugino,an,nonuni2,Baer,martin} in the gaugino sector and in the Higgs sector. Conclusions are given in Sec.(V).

\section{II. Analysis}

We will first carry out the analysis in the framework of mSUGRA and then 
extend the analysis to SUGRA models with non-universalities. We begin with a set of $\sim$50,000,000 
mSUGRA and non-universal SUGRA models passing radiative breaking of electroweak symmetry (REWSB).

The general constraints applied immediately include limits on sparticle masses from LEP~\cite{pdgrev}, 
on $\Omega h^2$ from WMAP~\cite{WMAP}, on $g_\mu -2$~\cite{Djouadi:2006be}, and constraints on 
B-physics~\cite{bphys} from FCNC data (this also includes the previous limit~\cite{Abazov:2010fs} on $\brbsmm$).
Specifically, we impose $0.0896 < \Omega_\chi h^2 < 0.1344$, %WMAP constraint
$\left(-11.4\times 10^{-10}\right)\leq \delta \left(g_{\mu}-2\right) \leq \left(9.4\times10^{-9}\right)$, %g_mu constraint
$\left(2.77\times 10^{-4} \right)\leq {\mathcal{B}r}\left(b\to s\gamma\right) \leq \left( 4.37\times 10^{-4}\right)$ %bsgnlo
(where this branching ratio has the NNLO correction~\cite{Misiak:2006zs}), % Misiak cite
${\mathcal{B}r}\left(B_{s}\to \mu^{+}\mu^-\right)\leq 4.3\times10^{-8}$, % old bsmumu constraint
and the sparticle mass limits
$m_h > 92.8~\GeV$, 
$m_{\sta} > 81.9~\GeV$, 
$m_{\cha} > 103.5~\rm GeV$, 
$m_{\ta} > 95.7~\GeV$, 
$m_{\ba} > 89~\GeV$, 
$m_{\ser} > 107~\GeV$, 
$m_{\smr} > 94~\GeV$, 
and $m_{\g} > 308~\GeV$.
The  relic density and the indirect constraints 
were calculated using {\tt micrOMEGAs}~\cite{micro}, 
and the sparticle mass spectrum was calculated using {\tt SuSpect}~\cite{SuSpect}.
In addition to the above we will also apply the LHC-7 and XENON-100 constraints when considering the
90\%~C.L. CDF signal region  as given in Eq.(\ref{3}), i.e.,   $4.6\times 10^{-9} < \brbsmm < 3.9 \times 10^{-8}$.

CMS and ATLAS have recently reported their first results for supersymmetry searches~\cite{cmsREACH,AtlasSUSY,atlas0lep} based on $35~\pb$ of data. The implications of these results 
have been considered for the  parameter space of SUGRA models 
in a number of works~\cite{Akula:2011zq,LHC7,Akula2}.
Here we will use the constraint arising from the $165~\pb$ of data from ATLAS. 
Similarly, the recent constraints from the CDMS~\cite{cdms} and XENON-100~\cite{xenon} experiments
have been analyzed also in a number of 
works~\cite{Akula2}. In this work we will explore simultaneously 
the ATLAS $165~\pb$ constraint, the constraint from XENON-100, as well as the new two-sided limit on $\brbsmm$, given in Eq.(\ref{3}) from CDF.

\begin{figure}[t!]
\begin{center}
\includegraphics[scale=.065]{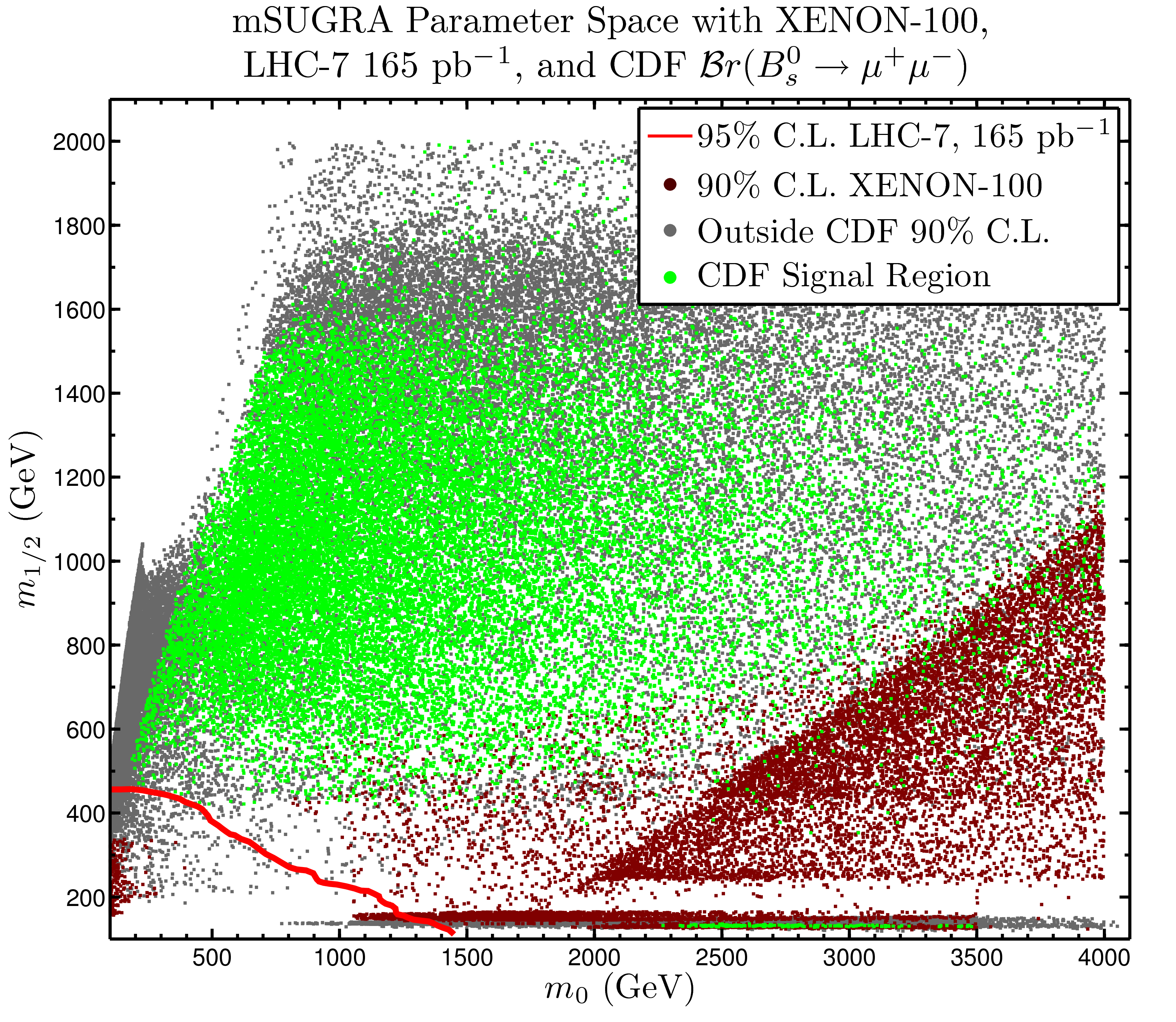}
\end{center}
\caption[]{ \label{bsmu}
The mSUGRA parameter space in the $m_0$ and $m_{1/2}$ plane allowing  for $A_0/m_0$ range $(-10,10)$ and
$\tan\beta$ range $(1, 60)$ in the initial scan.
The grey area is the region which satisfies all the previous experimental 
constraints prior to the new results from LHC-7, XENON-100 and the CDF 
$\brbsmm$ result. Specifically it has only the $\brbsmm <4.3\times 10^{-8}$. 
The red curve is the 95\%~C.L. ATLAS exclusion curve based on the 0~lepton search over
$165~\pb$ of integrated luminosity. 
In maroon, we show the models excluded by the new XENON-100 results. 
Finally, we highlight in lime green the unconstrained mSUGRA models within the 
90\%~C.L. CDF signal region  as given in Eq.(\ref{3}), i.e.,   $4.6\times 10^{-9} < \brbsmm < 3.9 \times 10^{-8}$ 
and these models correspond to values of $\tan\beta \in (28,58)$ and $A_0 \in (-3.0, 3.6) m_0$.
Relic density consistent  with WMAP, i.e., $0.0896 < \Omega_\chi h^2 < 0.1344$ is satisfied over the entire dotted region
exhibited in the figure above.  
}
\vspace*{-2mm}
\end{figure}

After applying the general constraints described above (i.e., all constraints listed above other than XENON-100, LHC-7, and CDF),
we give a survey of the mSUGRA parameter space in Fig.(\ref{bsmu}). Models excluded by XENON-100 are given in 
maroon; the LHC-7 constraint is readily visible as the models below the red 95\%~C.L. LHC-7 exclusion curve. We highlight  (in lime green)  the
models which are not constrained by any previous experiment and produce a consistent value of $\brbsmm$ in the
range $4.6\times 10^{-9} < \brbsmm < 3.9 \times 10^{-8}$.  In contrast, models
that satisfy the XENON-100 limit  and other constraints but fall outside of the new CDF signal region are shown in dark grey.
The models  below the LHC-7 curve mostly have the $\sta$ as the NLSP. We remark on the large absence
of models in the space that is presently  constrained by LHC-7~\cite{Akula:2011zq}, where the absence
of models below the LHC-7 limits arises due to previous indirect experimental constraints (see the previous paragraph for these constraints). Models failing the XENON-100 limit mostly lie on 
the hyperbolic branch/focus point (HB/FP) region~\cite{HB} where $\TeV$ size scalars co-exist with 
a small $\mu$. 
We note that $m_0$ up to 30~$\TeV$, for trilinear of comparable size,
 can still be
compatible with a sub-$\TeV$ $\mu$~\cite{Feldman:2011ud}. 
 Generic models with large $m_0$ would give a value of $\brbsmm$ close to the SM  prediction.

We see that the combined constraints have a sweeping effect on the Higgs pole region (see Ref.~1~of~\cite{LHC7}), where 
$m_0$ grows large along the low $m_{1/2}$ limit. Here, the correct relic density is produced by neutralino annihilations via $h^0$. And, 
the low masses of $\g$ and $\na$ give rise to potentially early discovery by XENON-100 and LHC-7. The XENON-100 results have already 
constrained a portion of this region, but it has left the very long strip seen in grey and green in Fig.(\ref{bsmu}). However,
the new CDF $\brbsmm$ limit leaves  only a $\rm few~\TeV$ window in $m_0$ in this region of signature space.

\begin{figure}[t!]
\begin{center}
\includegraphics[scale=.065]{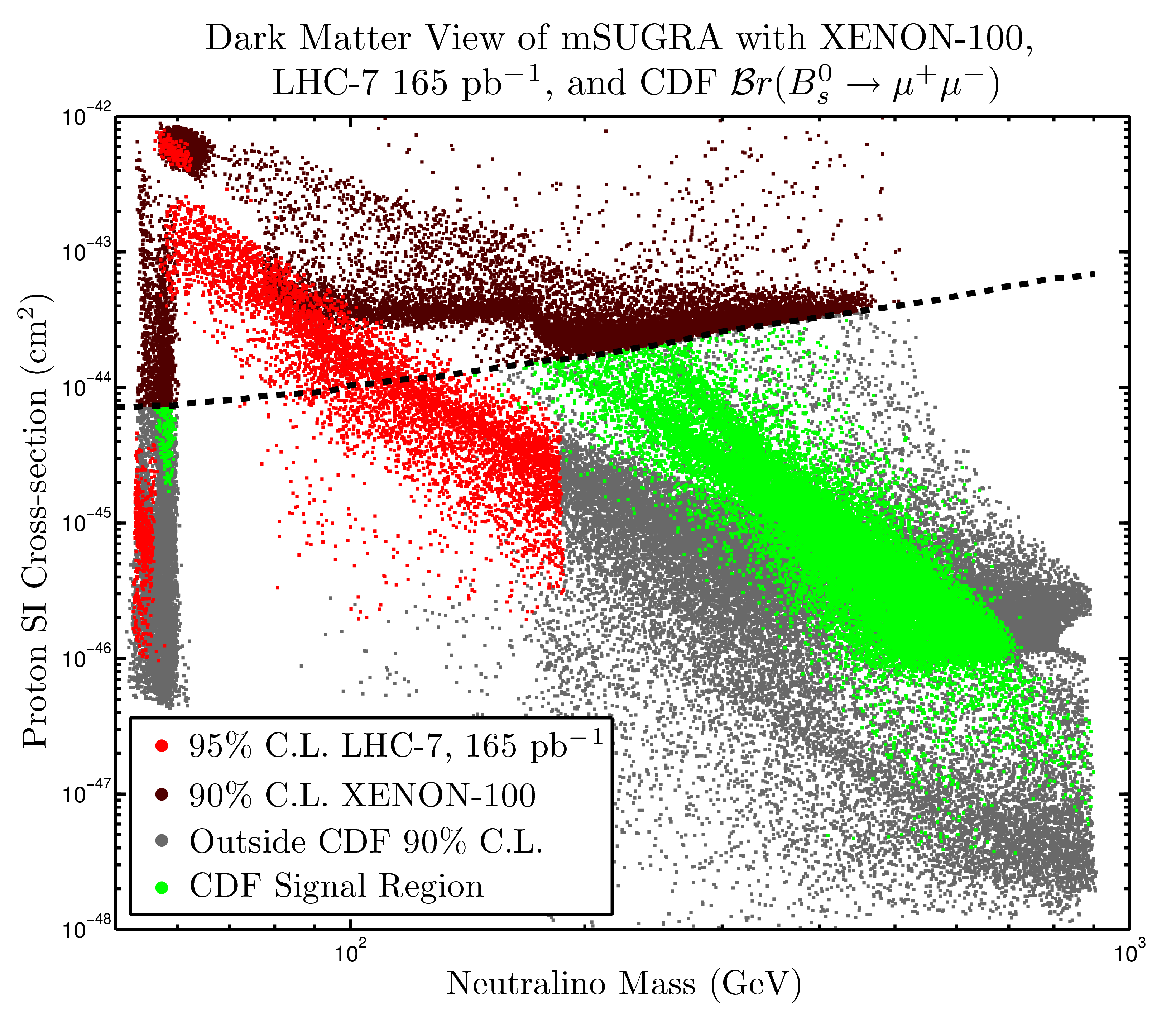}
\caption[]{ \label{xenon}
An exhibition of the spin independent neutralino-proton cross-section
against the neutralino mass. All models shown satisfy the general 
constraints discussed first in  Sec.(II) (see fig.(\ref{bsmu})). Models in maroon are excluded by the XENON-100 limit. Models in red are 
excluded by the ATLAS LHC-7 $165~\pb$~0~lepton search. In grey, we have the 
models satisfying both the LHC-7 and XENON-100, but outside the two-sided 
90\%~C.L. CDF $\brbsmm$ limit. Finally, in lime green we highlight the mSUGRA 
models passing all constraints and are within the new two sided bound on  $\brbsmm$
i.e. the signal region in the CDF data.
}
\vspace*{-2mm}
\end{center}
\end{figure}

 In Fig.(\ref{xenon}) we
 give a similar analysis for  the spin independent neutralino-proton 
 cross-section vs the neutralino mass. The grey, maroon and lime green regions have the same
 meaning as in Fig.(\ref{bsmu}) except that here, we explicitly color the models constrained by 
 LHC-7 in red. That is, in grey we have models not excluded by XENON-100 or LHC-7 but outside the 
 CDF signal region, and in maroon we have models failing the XENON-100 limits, 
 and we again highlight the models that fall within the 
 two-sided CDF $\brbsmm$  limit in lime green.

\begin{figure}[t!]
\begin{center}
\includegraphics[scale=.065]{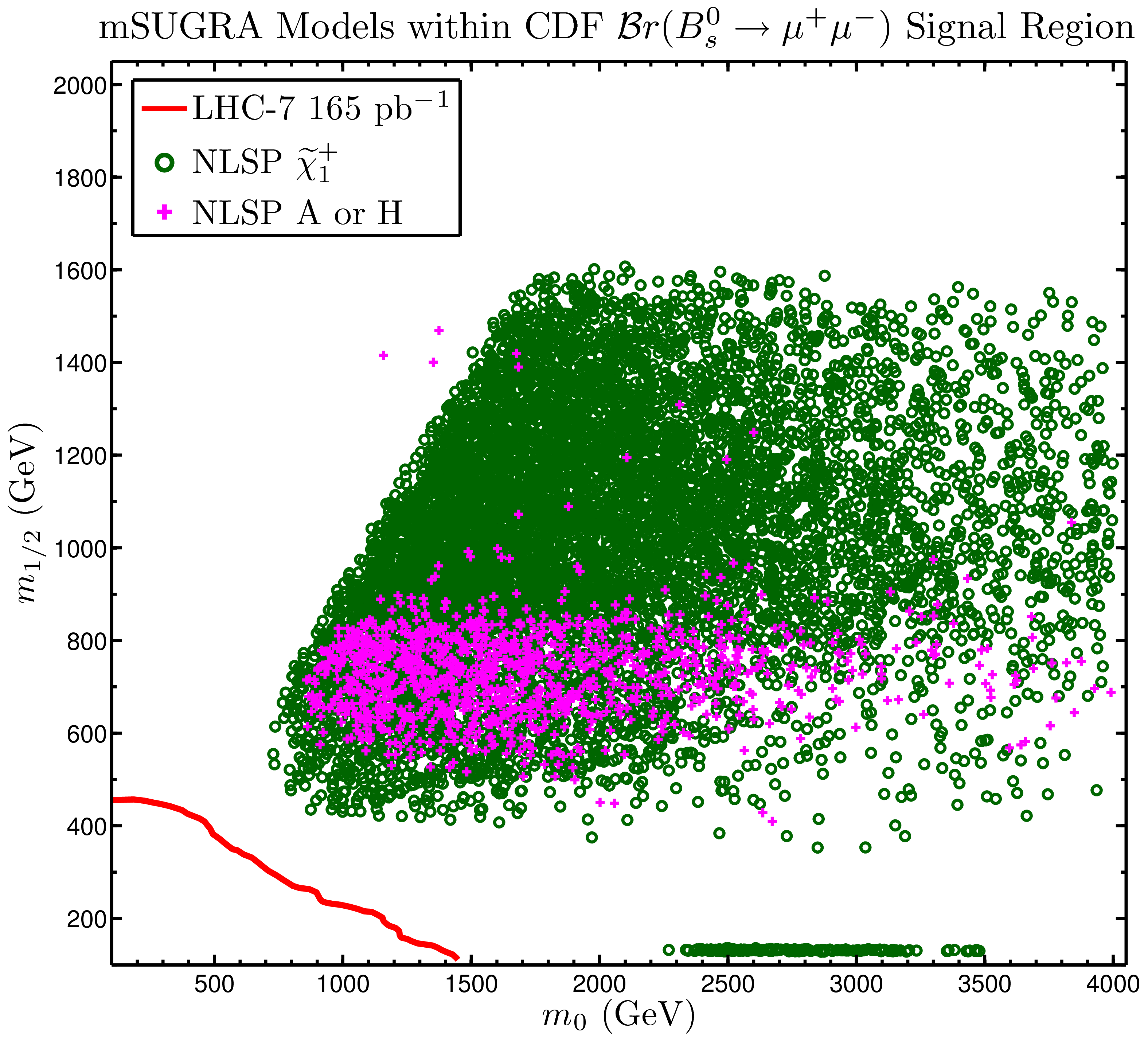}
\includegraphics[scale=.065]{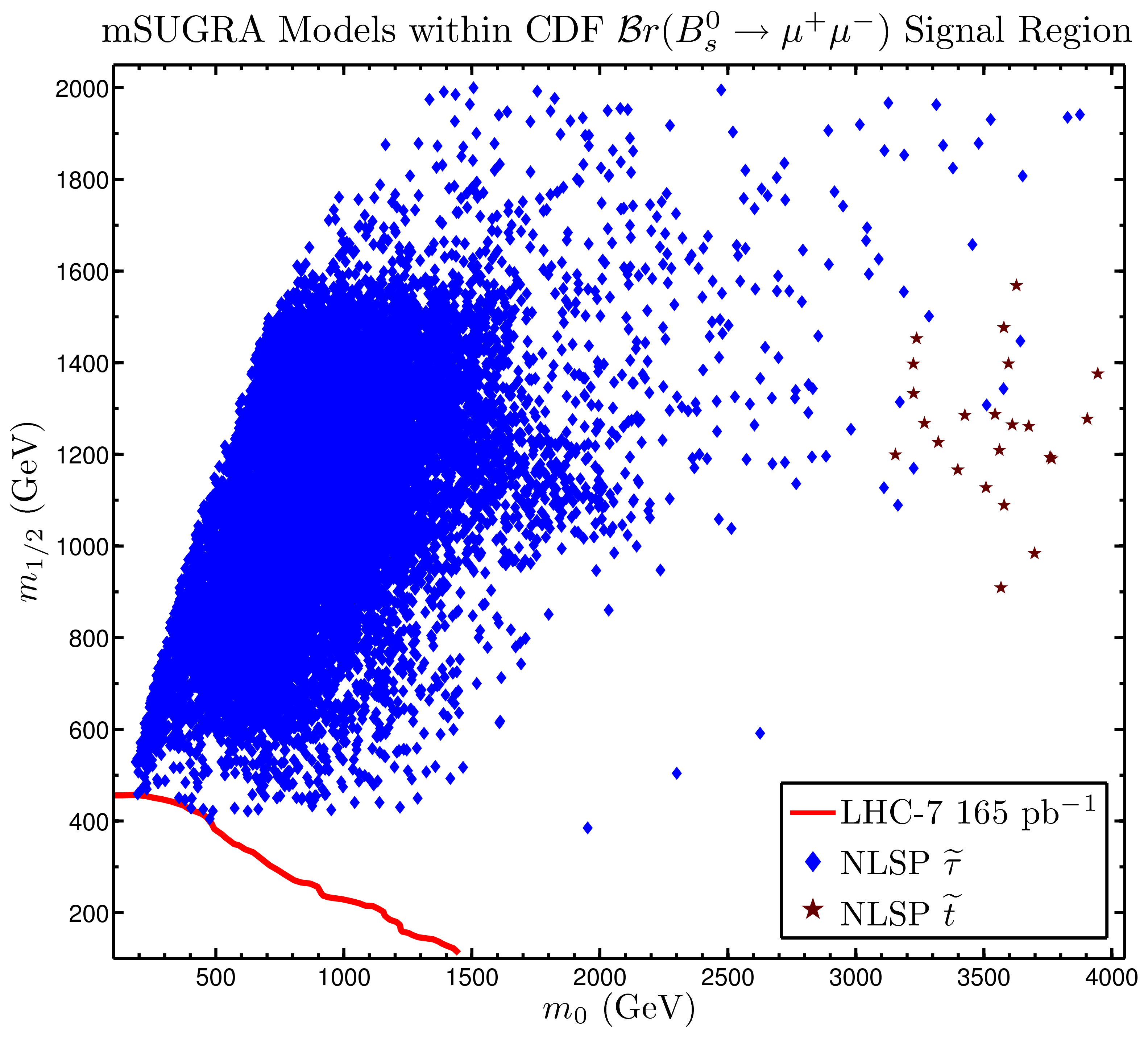}
\end{center}
\caption[]{\label{nlsp}
An exhibition of the allowed parameter space of mSUGRA satisfying all
constraints including the LHC-7, XENON-100 
producing a  $\brbsmm$ that can lie  within the CDF signal region.
 The models here are labeled by  the NLSP. The left panel shows
models with chargino and CP even or odd Higgs as NLSP. The right panel 
shows models with stau or stop as the NLSP. The stop NLSP models is depleted 
by the CDF result but can be seen in the upper right region of the right panel.
}
\vspace*{-2mm}
\end{figure}

 We now investigate the NLSPs allowed in the new CDF signal region (shown in lime green 
 in the Fig.(\ref{bsmu}) and Fig.(\ref{xenon}). Exhibited in Fig.(\ref{nlsp}), are the models
 passing all constraints, including LHC-7 as well as XENON-100, and within the CDF 90\%~C.L. two-sided 
 limit on $\brbsmm$. In an attempt to avoid illusion from significant overlapping of NLSP regions 
 we present in the left panel the chargino and CP even or odd Higgs NLSP models, and in the right panel, 
 we give the stau and stop NLSP models. The number of stop NLSP models is severely depleted by 
 the new CDF $\brbsmm$ limit, and can only be seen in the relatively high $m_0-m_{1/2}$ region. We 
 can see from the left panel, the remainder of the Higgs pole region, as well as the HB/FP region.

From our large pool of mSUGRA models passing REWSB, once all constraints have been applied, including 
the LHC-7 and XENON-100 limits, we have $\sim$25k models in CDF signal region. From this set, we 
discuss the range of masses remaining, in particular the scalar superpartners. We first note that our 
original set of models exhibited mass spectra close to the limits in~\cite{pdgrev} (given above).
The novelty of the two-sided CDF $\brbsmm$~result is that we should be able to observe upper bounds on 
sparticle masses, specifically scalar masses. This is because for large values of $m_0$, mSUGRA models 
will give $\brbsmm$~in accordance with the standard model. We indeed find that the models in the 
CDF signal region do not contain a stop with mass greater that $3.2~\TeV$ nor a stau with mass greater 
than $2.8~\TeV$, in sharp contrast to the largest masses from our original set. Also, we note that
while~\cite{pdgrev} gives a lower limit for the stop mass at $95.7~\GeV$, the lowest stop mass that we 
observe in this region is $450~\GeV$.
 
\section{III. LHC signatures}

For the LHC-7 analysis, we use the simulated SM background of~\cite{Peim} 
  which was generated with {\tt MadGraph 4.4}~\cite{madgraph} for  parton level processes, {\tt Pythia~6.4}~\cite{pythia} for  hadronization and {\tt PGS-4}~\cite{pgs} for detector simulation. An MLM matching algorithm with a $k_{T}$ jet clustering
scheme was used to prevent double counting of final states.
A more thorough discussion on the details of this background can be found in~\cite{Peim}
(see also~\cite{Lessa} for discussions on SM background for $2\to N$ processes).  For LHC-7 event analysis we used a modified version of {\tt Parvicursor}~\cite{baris}.
Further the sparticle decays of the models are calculated using {\tt SUSY-HIT}~\cite{susyhit} before being simulated at the LHC.
We note in passing, that there are cases where the variations in the gluino mass are observed between {\tt SoftSUSY}~\cite{softsusy} and {\tt SuSpect}~\cite{SuSpect} at the level of  $\sim (5-10)\%$, for the same soft breaking inputs at the high scale due to different approaches in the simulations of the sparticle spectra.

A portion of the models in the CDF signal region are close to being visible to both the LHC-7 and dark matter direct detection experiments.
To illustrate this point, we begin with models that have neutralino mass and neutralino-proton spin independent cross-section close to the XENON-100 limit,
and the present LHC-7 limit. We investigate signatures that could lead to visibility at LHC-7 with $5~\fb$ of 
integrated luminosity using two benchmark models. 
The benchmarks are chosen from our large set of models that  provide a prediction for  $\brbsmm$ that fall within 
the CDF signal region.  as discussed in Sec.(II) including the most recently 
reported LHC bounds at $165~\pb$~\cite{atlas165pb}. One of these is a Higgs pole model, which will be 
referred to as Model~1, and the other is a stau NLSP model, which will be referred to as Model~2.  For 
each model, the  mSUGRA parameters and a few selected masses are shown in Table~\ref{benchmass}. 
Further, we give the neutralino-proton spin independent cross-sections which indicate that these 
two benchmarks will soon be in the sensitive region of dark matter direct detection experiments.

\begin{table}[t!]
\begin{center}
\begin{tabular}{|l|c|c|c|c|c|c|}
\hline
 &$m_{\na}$ & $m_{\cha}$ & $m_{\g}$ & $m_{\sta}$ & $m_{h}$ & $m_{A}\simeq m_{H}$  \tnhl
Model~1 & 58 & 116 & 442 & 1709 & 118 & 565 \tnhl
Model~2 & 193 & 359 & 1086 & 201 & 114 & 562 \tnhl
\end{tabular}
\vskip 0.2cm
\begin{tabular}{|l|c|c|c|c|c|c|}
\hline
 &  $m_0$&$m_{1/2}$& $A_0$&$\tan\beta$ & $\brbsmm$ &  $ \sigma_{\na p}^{\rm SI}~({\rm cm}^{2})$ \tnhl
Model~1 & 2455 & 131 & -2069 & 49.3 & $9.34\times 10^{-9}$ & $5.4\times 10^{-45}$ \tnhl
Model~2 & 213 & 471 & 667 & 30.5 & $4.98\times 10^{-9}$ & $5.6\times 10^{ -45}$\tnhl
\end{tabular}
\caption{\label{benchmass} Exhibition of the selected sparticle masses of two benchmarks as well as the model parameters.  These benchmarks were generated in {\tt micrOMEGAs}~\cite{micro} using {\tt SuSpect}~\cite{SuSpect} for the RGEs with $m_{\rm top}^{\rm pole}=173.1~\GeV$ and $\sgn{\mu}>0$.  All masses are given in units of GeV.  These models pass all constraints including LHC-7, XENON-100 and the CDF result on $\brbsmm$.}
\end{center}
\end{table}

We follow similar preselection requirements that ATLAS reports in~\cite{atlasTDR}. Jet 
candidates must  have $p_{T}>20~\GeV$ and $\left|\eta\right|<4.9$ and electron candidates must 
have $p_{T}>10~\GeV$ and $\left|\eta\right|<2.47$.  Events are vetoed if a ``medium" electron~\cite{atlasTDR} 
is in the electromagnetic calorimeter transition region, $1.37<\left|\eta\right|<1.52$.  Muon candidates must 
have $p_{T}>10~\GeV$ and $\left|\eta\right|<2.4$. 
Further, jet candidates are discarded if they are within $\Delta R=\sqrt{(\Delta \eta)^2+(\Delta \phi)^2}=0.2$ 
of an electron or muon.   Photon candidates must have $p_{T}>10~\GeV$ and $\left|\eta\right|<2.37$.  For the analysis, the (reconstructed) missing energy, $\met$, for an event is the 
negated vector sum of the $p_T$ of all the jet and lepton candidates.  As a preselection criterion  we require that events have $\met>60~\GeV$.

For the remainder of the event analysis, all objects are ordered by momentum and when referring to different cuts we 
define cuts on ``selected" objects, i.e. jets or leptons, to mean that the ``selected" objects candidate has bare minimum number of these objects.  In an event selected lepton candidates are required to have $p_{T}>20~\GeV$ and  all selected jet candidates must have $\left|\eta\right|<2.5$ and $p_{T}>30~\GeV$.  
Further, events are rejected if the missing energy points along the same direction as any of the selected jets., i.e. we require $\Delta \phi \left(j_{i},\met\right)>0.4$, 
where $i$ is over the ``selected" jets.  Events are required to have no photon candidates and $\met>150~\GeV$. 
 To prevent contamination from fake missing energy, which could arise from a visible object failing selection criteria, we require that the reconstructed missing energy and the missing energy in the calorimeter, $\met^{\rm cal}$, satisfy 
 $R_{miss}=\met/\met^{\rm cal}<1.25$.  We also define the effective mass, $\meff$, and $H_{T}$ to be
  \begin{eqnarray}
 \meff\left(n_{j},n_{\ell}\right)&\equiv&\displaystyle\sum_{i=1}^{n_{j}}p_{T}\left(j_i\right) +\displaystyle\sum_{k=1}^{n_{\ell}}p_{T}\left(\ell_k\right)+\met\\ H_{T}\left(n_{j},n_{\ell}\right)&\equiv &\meff\left(n_{j},n_{\ell}\right)-\met~,
 \end{eqnarray}
where $n_{j}$ ($n_{\ell}$) are the number of selected jets (leptons).  Note in the case that $n_{j}=0$ or $n_{\ell}=0$ that the sum over the given object is omitted.  
Taking into account the above definitions and criteria we design three signals to probe the large $\g\g$ and $\cha\nb$ production~(Ref.~1~of~\cite{LHC7}) of the Higgs pole model (Model~1) as well as the large tau production in the stau model (Model~2).  This leads us to investigate the following signals for the benchmarks:
\begin{eqnarray*}
{\rm Signal~1:}&~~~&n\left(\ell\right)=0,~n(j)\geq 4,~p_{T}\left(j_1\right)>100~\GeV,~p_{T}\left(j_4\right)\geq 40~\GeV\\
{\rm Signal~2:}&~~~& n\left(\ell\right)\geq 2,~n\left(j\right)\geq 2,~p_{T}\left(j_1\right)\geq 100~\GeV,~\meff(2,2)\geq 400\\
{\rm Signal~3:}&~~~& n\left(\ell\right)\geq 2,~n\left(j\right)\geq 2,~p_{T}\left(j_1\right)\geq 100~\GeV,~\met>0.2\times\meff(2,2),~H_{T}(2,2)\geq 400~,
\end{eqnarray*}
and we display significance for each model, $S/\sqrt{B}$, in Table~\ref{sigTab} at $5~\fb$ of integrated luminosity.  This is where the number of signal (background) events is given by $S~(B)$.  
Despite the large significance in Signal~1 for Model~1 our simulations show that the effective cross sections for the signal channels in the ATLAS~0~lepton searches at $35~\pb$~\cite{atlas0lep} and at $165~\pb$~\cite{atlas165pb} are consistent with their reported bounds.  Explicitly we find in the  ATLAS~0~lepton search at $165~\pb$~\cite{atlas165pb} that Model~1 gives the effective cross-sections $11~{\rm fb}$, $18~{\rm fb}$ and $21~{\rm fb}$ in the three signal regions compared to the ATLAS reported $95\%$~C.L. upper bounds of $35~{\rm fb}$, $30~{\rm fb}$ and $35~{\rm fb}$.

\begin{table}[t!]
\begin{center}
\begin{tabular}{|l|c|c|c|}
\hline
& ${\rm Signal~1}$& ${\rm Signal~2}$& ${\rm Signal~3}$\tnhl
Model~1 & 23.3 & 4.44 & 3.29 \tnhl
Model~2 & 0.96 & 0.27 & 3.97 \tnhl
\end{tabular}
\caption{\label{sigTab} Exhibition of the significance for each model of Table~\ref{benchmass} at LHC-7 for an integrated luminosity of $5~\fb$ of data for the cuts stated in the text.   Under these designed cuts these models can be probed in the next round of data, even the Higgs pole models where $m_0$ is large.  These models are consistent with the limits reported by the  ATLAS~0~lepton searches at $35~\pb$~\cite{atlas0lep} and at $165~\pb$~\cite{atlas165pb}. }
\end{center}
\end{table}

As can be easily seen from the left panel of Fig.(\ref{nlsp}), there remains many models lying
on the Higgs pole (discussed above) that pass all constraints, including XENON-100, LHC-7 and CDF $\brbsmm$.
As recently pointed out, and studied in detail, the Higgs-pole models within the minimal
unified model of soft breaking leads to a spectrum with a rather low mass gluino whose
mass range is constrained to be about $400~\GeV$ to $600~\GeV$  and if present, its signatures
in jets, missing energy and dileptons must manifest at the LHC~as seen in Ref.~1~of~\cite{LHC7}. 

\begin{figure}[t!]
\begin{center}
\includegraphics[scale=.40]{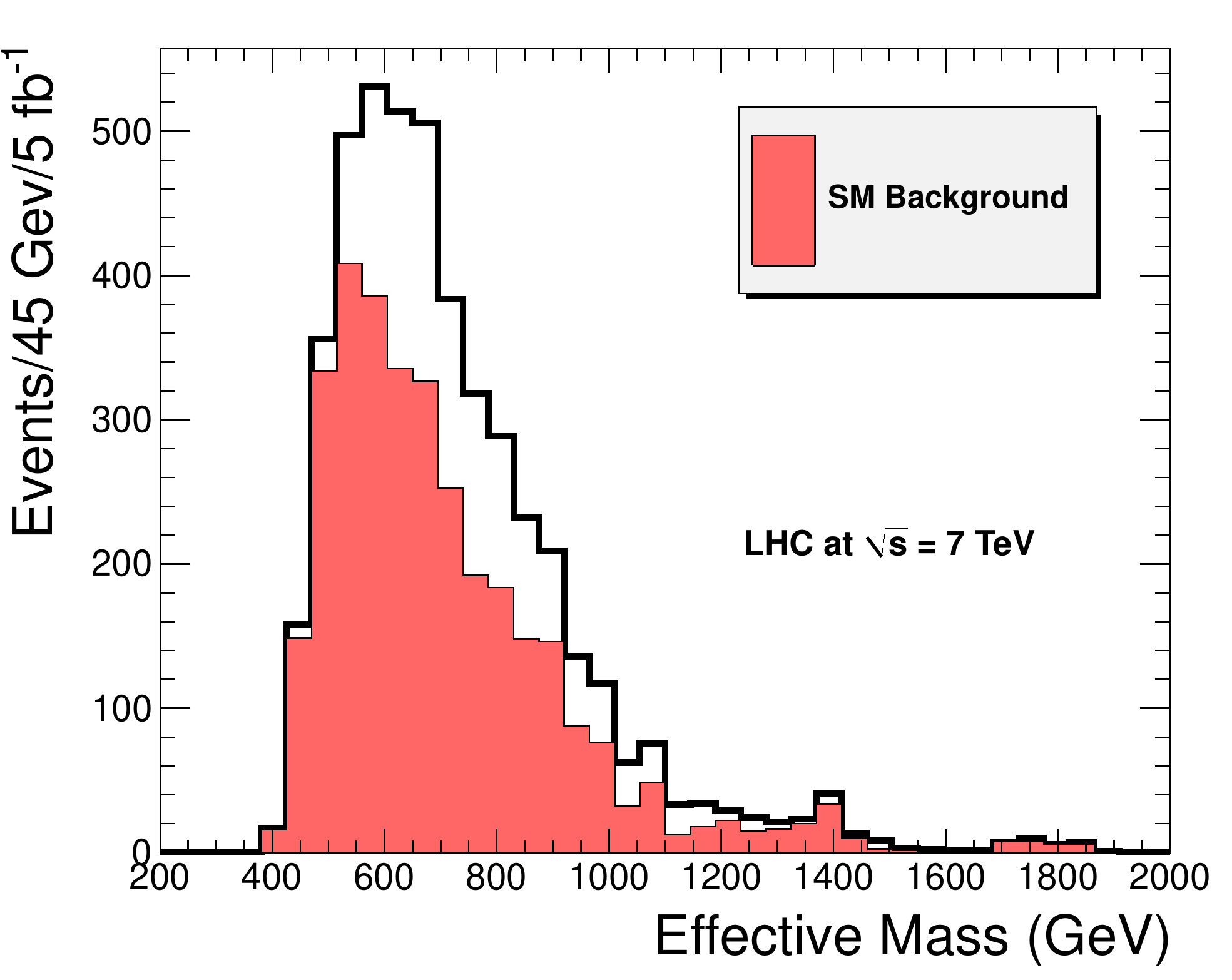}
\includegraphics[scale=.35]{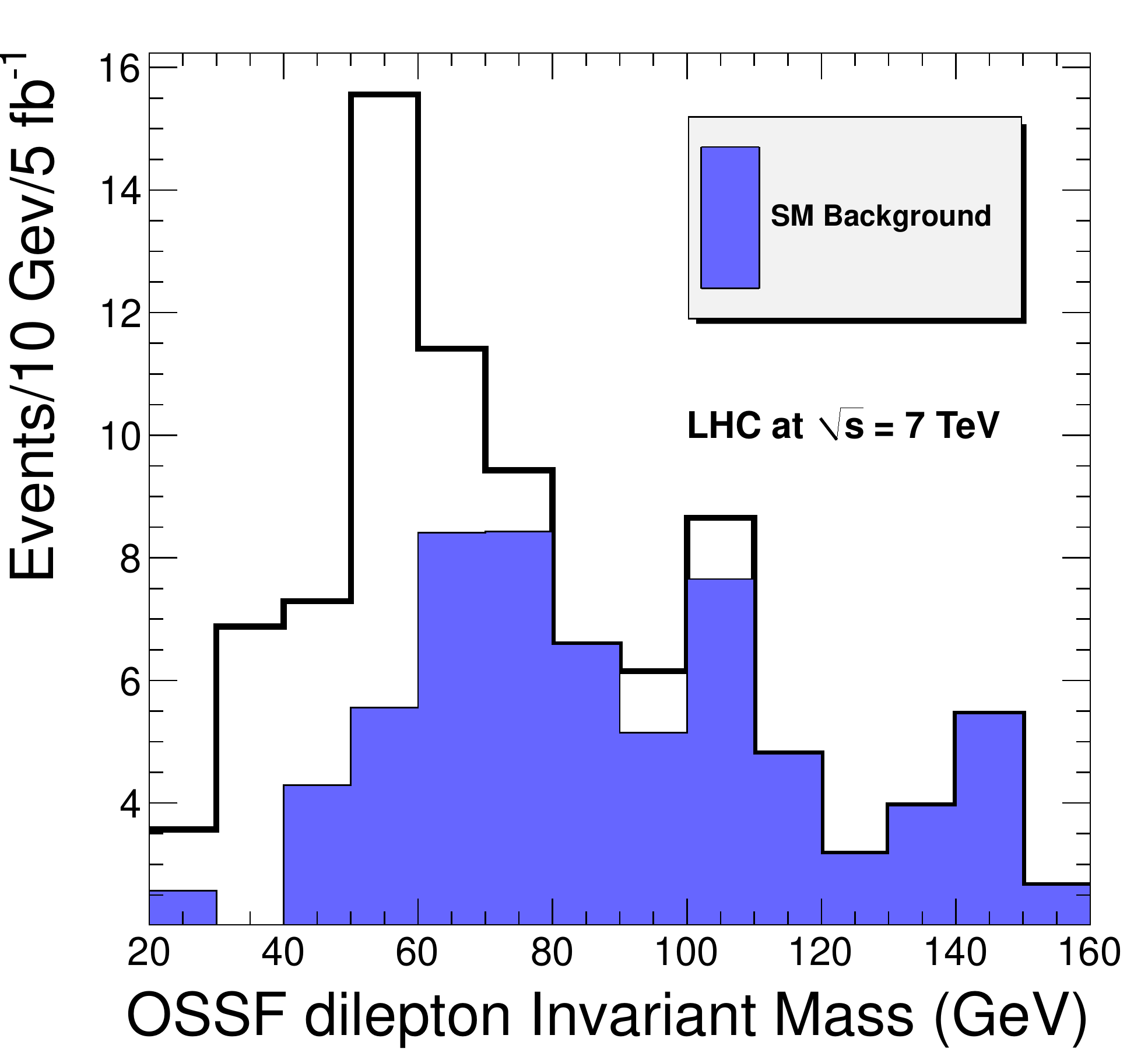}
\caption{\label{sigLHC} 
Signature analysis of Model 1 consistent with all constraints including constraints from LHC-7, XENON-100, and 
the CDF result on $\brbsmm$. Left: A distribution of the effective mass, $\meff(4,0)$, in 45~GeV bins for 5~fb$^{-1}$ of
integrated luminosity at LHC with $\sqrt{s}=7$~TeV using  ${\rm Signal}~1$ described
in the text. Exhibited is the signal plus SM background compared to the 
 SM background.
Right: An exhibition of the  number of opposite sign same flavor  (OSSF) dileptonic events for the same model point as in the left panel in 10~GeV bins
at 5~fb$^{-1}$ of integrated luminosity using ${\rm Signal}~2$ as described in the text.  The dark region is 
the SM background and the white region is the sum of the background plus signal.  }
\end{center}
\end{figure}

To illustrate this point we have displayed the effective mass, $\meff(4,0)$, and the opposite sign same flavor (OSSF) dilepton distributions for Model~1 under Signal~1 and Signal~2 in Fig.(\ref{sigLHC}).  
Since the Higgs pole model mass spectrum corresponds
to heavy squarks, the gluino mass can remain light in accordance  with the LHC limits once one can establish a relationship between the effective mass
and the  gluino mass as shown in~Ref.~1~of~\cite{LHC7}.
In the left panel a clear excess can be seen with a peak at about $630~\GeV$, which is consistent with the gluino mass relationship shown in~Ref.~1~of~\cite{LHC7}.  
The Higgs pole region also corresponds to a neutralino mass near $50~\GeV$ and
it populates a region in the neutralino mass-$\sigma_{\rm SI}$ plane where the XENON detector is most sensitive~\cite{xenon,Feldman:2011me}.  
Using the scaling properties among the gaugino masses, i.e. $m_{\g}\simeq (7-8)m_{\na}$, one can obtain a neutralino mass from the OSSF dilepton invariant mass distribution.  
As seen in the right panel we begin to see an edge forming with $5~\fb$ of integrated luminosity at  $55~\GeV$, which is in agreement with the expected sparticle mass difference.

%%%%%%%

\section{IV. Non-universalities in the gaugino and Higgs sectors}

Our analysis thus far has been in the framework of using universal boundary 
conditions for the soft parameters at the GUT scale. However, the nature of 
physics at the Planck scale is not  fully understood and thus one may consider
boundary conditions at the unification scale which are non-universal, but consistent
with  REWSB, FCNC constraints, and the other relevant constraints discussed in Sec.(II). For the
gaugino sector we assume non-universalities in the gaugino masses so that 
$\tilde m_{i} = m_{1/2} (1+ \delta^G_i), i=1,2,3$ where $\delta^G_i$ parameterize
the deviations from non-universality and $i=1,2,3$ correspond to the $U(1)$, $SU(2)$
and $SU(3)$ elements of the standard model gauge group. The Higgs sector too may
have non-universalities so that the soft Higgs boson masses at the GUT scale
may be parameterized as follows: $m_{H_u} = m_0(1 + \delta_{H_u})$, and
$m_{H_d} = m_0(1 + \delta_{H_d})$, where $H_u$ gives mass to the up quarks and
$H_d$ gives mass to the down quarks and the leptons.

\begin{figure}[t!]
  \begin{center}
    \includegraphics[scale=.065]{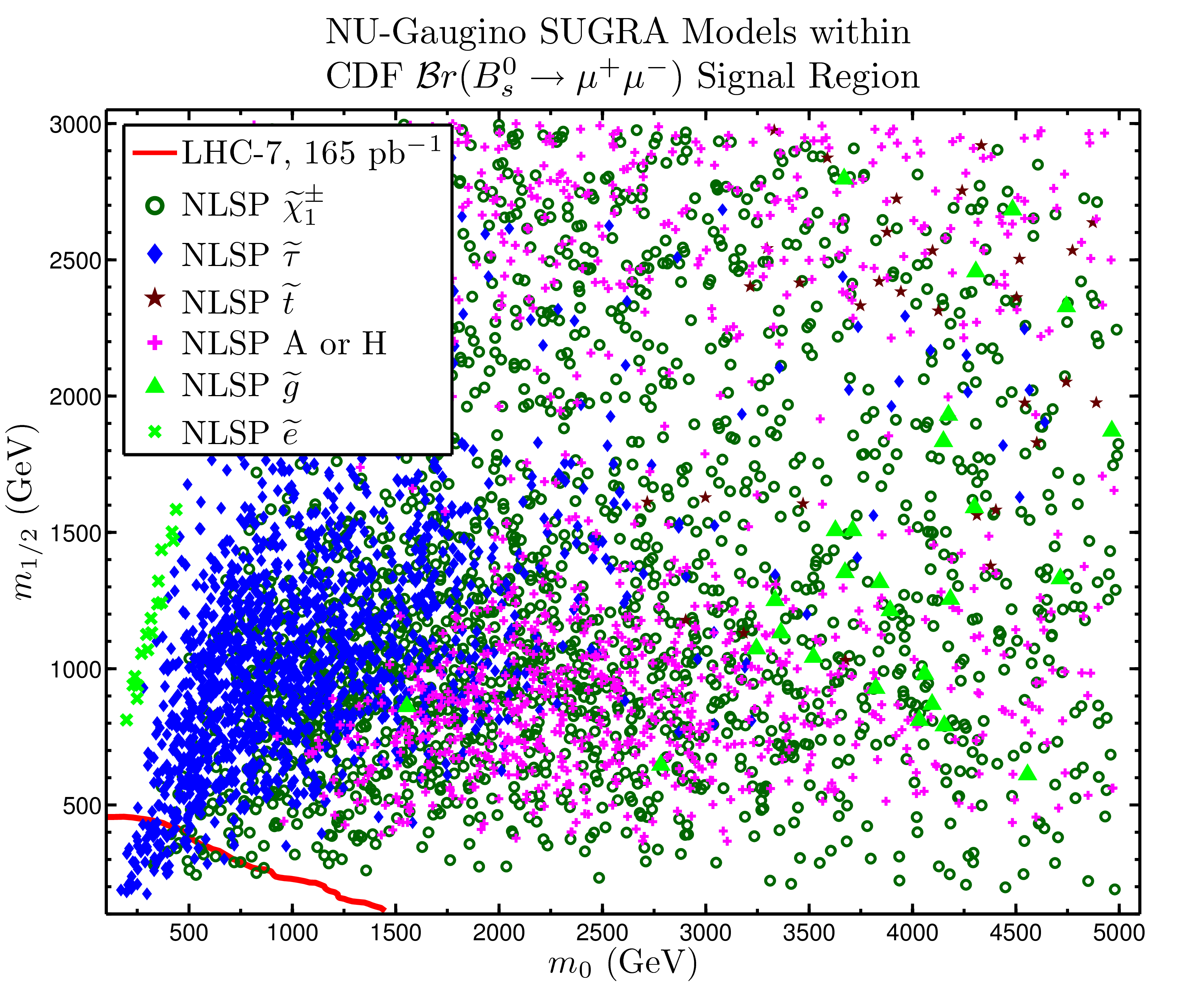}
    \includegraphics[scale=.065]{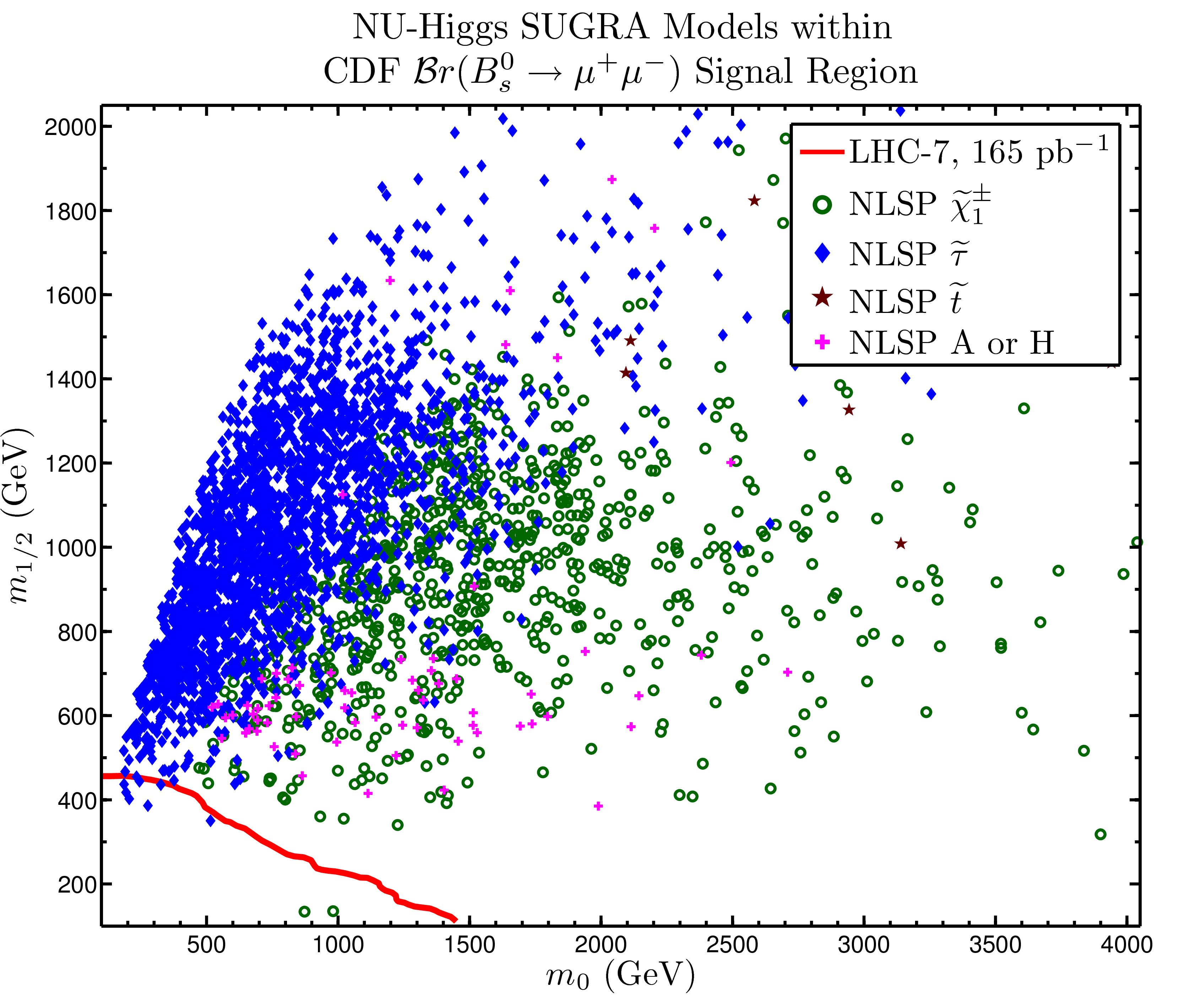}
    \vspace{-2mm}
    \caption{\label{non-uni}
      Exhibited here are cases of SUGRA with non-universal boundary conditions on 
      the gaugino sector (left) and the Higgs sector (right)  with initial scan allowing the  $A_0/m_0$ range $(-10,10)$ and
       $\tan\beta$ range $(1, 60)$. In each case, the models 
      presented satisfy limits from LEP, WMAP, FCNC data, and XENON-100, as well as 
      the new CDF 90\%~C.L. $\brbsmm$~double-sided limit. We readily see that the introduction 
      of non-universalities in the gaugino sector presents models with new NLSP cases 
      from mSUGRA, as we see models with $\g$ and $\ser$ NLSPs.
    }
  \end{center}
\end{figure}

An analysis of the CDF constraint  of Eq.(\ref{3}) including non-universalities is given
in Fig.(\ref{non-uni}). The left panel of Fig.(\ref{non-uni}) gives the analysis for the case
of gaugino mass non-universalities which has all the constraints including the LHC-7, XENON-100
and CDF limits constraints of Eq.(\ref{3}). A comparison of Fig.(\ref{non-uni}) with Fig.(\ref{nlsp})
shows that some areas depleted for the mSUGRA case are now re-populated. Further, one finds
that there arise new NSLPs in the landscape which are $\ser$ and $\g$ not present
for the mSUGRA case (see \cite{Feldman:2009zc} for a dedicated study of the $\g$NLSP class
of models in SUGRA,  first uncovered in \cite{Feldman:2007fq}). Further as was previously shown in Ref.~1~of~\cite{Akula2}, the 
region excluded by the LHC-7 data for mSUGRA is repopulated.
The analysis including non-universalities in the Higgs sector is exhibited in 
the right panel of Fig.(\ref{non-uni}). Here the differences from  the mSUGRA case
are less dramatic. One does not find new NSLPs. Also repopulation of the depleted 
area is relatively smaller compared to gaugino universality case and the repopulated
region does not extend into the area excluded by LHC-7.

\section{V. Conclusion}
The recent CDF results  indicate the first observation of a $\bsmm$
signal. In this work we have carried out an analysis under the combined constraints
of the recent LHC-7 data, the constraints from the 2011 XENON-100 data on the 
neutralino-proton spin independent cross-section and the CDF two-sided constraint 
on $\brbsmm$. The analysis reveals that within the mSUGRA parameter 
space remaining after all the constraints are imposed 
the NLSP is mostly either chargino, stau, CP even or odd Higgs boson, or more 
infrequently, a stop.
The limits on the stop and stau masses arising solely from the CDF 90\% C.L. limit on $\brbsmm$
is also exhibited. It is pointed out that the CDF result imposes upper limit
on sparticle masses.
 An analysis of signatures of benchmarks which are accessible 
at the LHC-7 is also carried out. The CDF results, if supported by further data
by LHCb as well as D0, would be a harbinger for the early discovery of sparticles at the LHC.
\\[12pt]
\noindent

After this work was done, a paper on the analysis of  $\bsmm$
appeared in~\cite{Dutta:2011bk} which has some overlap with this work.
\\[12pt]
\noindent 

{\it Acknowledgements:} 
We thank Darien Wood for a discussion. DF thanks the CERN Theory Group 
for the hospitality extended to him during the period when part of this work
was done.
 This research is  
supported in part by grants  DE-FG02-95ER40899,   PHY-0757959,  
and by  TeraGrid  grant 
TG-PHY110015.\\
  
%\clearpage

\end{document}